\documentclass[conference]{IEEEtran}

\usepackage{graphicx}
\usepackage{cite}
\usepackage[cmex10]{amsmath}
\usepackage{threeparttable}
\usepackage{amsmath}
\usepackage{amssymb}
\usepackage{diagbox}

\def\BibTeX{{\rm B\kern-.05em{\sc i\kern-.025em b}\kern-.08em
    T\kern-.1667em\lower.7ex\hbox{E}\kern-.125emX}}
\ifCLASSINFOpdf
\else
\fi
\begin{document}
%
\title{A 5.16Gbps decoder ASIC for Polar Code\\in 16nm FinFET}

\author{\IEEEauthorblockN{Xiaocheng Liu, Qifan Zhang, Pengcheng Qiu, Jiajie Tong, Huazi Zhang, Changyong Zhao, Jun Wang}
\IEEEauthorblockA{Huawei Technologies Co. Ltd.\\
Email: \{liuxiaocheng, Qifan.Zhang, qiupengcheng, justin.wangjun\}@huawei.com}}


%


\maketitle

\begin{abstract}
Polar codes has been selected as 5G standard. However, only a couple of ASIC featuring decoders are fabricated, and none of them support list size $L>4$ and code length $N>1024$.
This paper presents an ASIC implementation of three decoders for polar code: successive cancellation (SC) decoder, flexible decoder and ultra-reliable decoder. These decoders are all SC based decoder, supporting list size up to $1, 8 , 32$ and code length up to $2^{15}, 2^{14}, 2^{11}$ respectively.
 This chip is fabricated in a 16nm TSMC FinFET technology, and can be clocked at $1$ Ghz. Optimization techniques are proposed and employed to increase throughput. Experiment result shows that the throughput can achieve up to 5.16Gbps. Compared with fabricated AISC decoder and synthesized decoder in literature, the flexible decoder achieves higher area efficiency.
\end{abstract}



%
\IEEEpeerreviewmaketitle

\section{Introduction}
Polar codes, proposed by Arikan\cite{Arikan}, has been selected as the 5G standard. Although Polar codes with successive-cancellation (SC) decoding is proved to achieve channel capacity in the asymptotic sense, its error-correction  performance is inferior to that of low-density parity-check (LDPC) or Turbo codes at short or moderate lengths.
SC list (SCL) decoding, regarded as the most efficient decoding algorithm of polar codes, improves the error-correction performance but suffers from low latency and low throughput due to the serial nature of SC.
Much effort has been made to optimize the decoding of  Polar codes \cite{SSC,parallel_Architecture,6920050,7001058,7337462,7114328,goodbit,fast_and_flexible}. However, most works lack ASIC implementation and thus bear less practical relevance.

A couple of ASIC featuring decoders are fabricated in \cite{SC_ASIC,6858413,polarbear}. The chip presented in \cite{SC_ASIC} implements the SC decoding algorithm; The chip presented in \cite{6858413} implements the belief-propagation decoding algorithm. Both of them suffer from mediocre error-correction performance. The chip presented in \cite{polarbear} implements SCL decoding, but constrains the largest list size $L_{max}=4$ and largest code length $N_{max}=1024$, which limits its application scope.
\subsection{Motivation and Contribution}
This work is motivated by the desire to provide ASIC decoder to support polar codes research and speed up prototype building of 5G communication systems.
The ASIC decoder should have low latency and high error-correction performance, and support a wide range of list sizes and code lengths.
To satisfy all the desired properties, we integrated three decoders in one chip:  SC decoder, flexible decoder and ultra-reliable decoder.
\begin{itemize}
\item SC decoder is designed for low latency and long code length with $N=2^{15}$;
\item Flexible decoder is a SCL decoder with $N_{max}=2^{14}$ and $L_{max}=8$. The list size of the flexible decoder can be configured during runtime;
\item Ultra-reliable decoder is also a SCL decoder, designed for ultra-reliable scene with largest code length $N_{max}=2^{11}$ and list size $L=32$.
\end{itemize}
All the decoders support any code rate. This is the first ASIC implemented SCL decoder supporting $L>4$ and $N>1024$.
\par To improve throughput, several optimization techniques are proposed. We propose a new internal log-likelihood ratio (LLR) messages storage method which can reduce $86\%$ of the internal LLR memory. A serial list processing architecture is proposed to avoid the crossbar of LLR. This can reduce resource and improve timing performance. To improve utilization ratio of processing element (PE), we propose to decode two packages simultaneously, which can improve throughput by $54\%$.
We also recovery decoded bit from partial-sum to save memory.
\subsection{Outline}
\par The rest of this paper is organized as follows. Section \uppercase\expandafter{\romannumeral2} gives a brief review of polar codes and SC-base decoding algorithms. The proposed ASIC architectures and optimization techniques are presented in Section \uppercase\expandafter{\romannumeral3}.
Section \uppercase\expandafter{\romannumeral4} presents the implementation results and comparison with state-of-the-art works.
Section \uppercase\expandafter{\romannumeral5} concludes the paper.

\section{Polar Code}
An $(N,k)$ polar code has a code length $N$ and $k$ information bits. The code rate $R$ is defined by $R=k/N$. The information bits are assigned to the $k$ most reliable sub-channels, and the remaining sub-channels are assigned by pre-defined value, typical zero---called frozen bits.
The encoding of Polar code can be defined as $c=uG$, where $u$ is the source vector, $G$ is the generator matrix, defined as $G \triangleq F ^ {\otimes n}$, where
$F=\bigl[ \begin{smallmatrix} 1 & 0 \\ 1 & 1 \end{smallmatrix} \bigr]$
is the kernel, $^\otimes$ denotes Kronecker power, and $n=log_2N$.
\subsection{SC-based Decoders}
\begin{figure}
\centering
\includegraphics[width=2.65in]{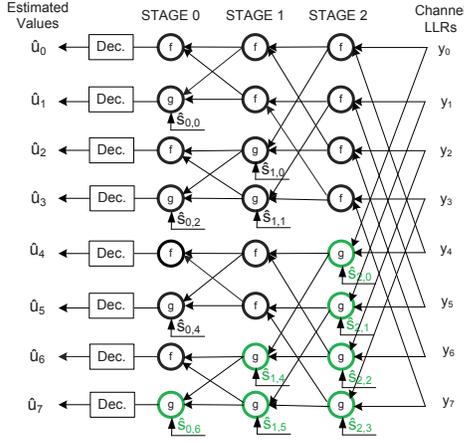} 
\caption{decoding graph.}
\label{decoding_graph}
\end{figure}

The decoding graph of SC decoder is shown in Fig. \ref{decoding_graph}. The soft values propagate from right to left and the hard bits propagate from left to right. The information vector $u$ is decoded sequentially from top to bottom. A hardware-friendly version of soft value updating is carried out in log-likelihood ratio (LLR) domain. Two incoming LLRs ($L_{in1}$ and $L_{in2}$) are combined to produce $L_{out}$ with the following f-function
\begin{equation}
L_{out}=sign(L_{in1} \cdot L_{in2}) \cdot min(|L_{in1}| , |L_{in2}|) .
\end{equation}
or g-function
\begin{equation}
L_{out}=L_{in1} + (-1)^{\hat{s}} \cdot L_{in2} ,
\end{equation}
where $\hat{s}$ is called partial sum (PS). For an SCL decoder, the decoding process is similar to SC decoder except that it keeps $L$ paths. When making hard decision for each bit, L paths split into 2L paths, and the ones with smallest path metric (PM) are kept. For list size $l$ and bit $u_i$, the LLR of stage $0$ is denoted as $L_{0,i}^l$ and its hard decision is denoted as $\beta_{0,i}^l$.
The PM updates according to
\begin{equation}
PM_i^l=  \left\{ \begin{array}{lll}
         PM_{i-1}^l,  &  \mbox{if} \   u_i^l = \beta_{0,i}^l \\
         PM_{i-1}^l+|L_{0,i}^l|,  &  \mbox{otherwise}
         \end{array}\right.
\end{equation}
After all bits are decoded, the path with the smallest PM is selected as the decoding output.
To further improve error correction performance, concatenated polar code is proposed.
For cyclic redundancy check (CRC) aided SCL (CA-SCL) \cite{CA_polar}, the most reliable path that passes the CRC is selected as the decoding output. For parity-check SCL (PC-SCL) \cite{PC-polar}, each parity bit is decided by its parity function rather than by the LLR.
\section{Architecture}
\begin{figure}
\centering
\includegraphics[width=2.75in]{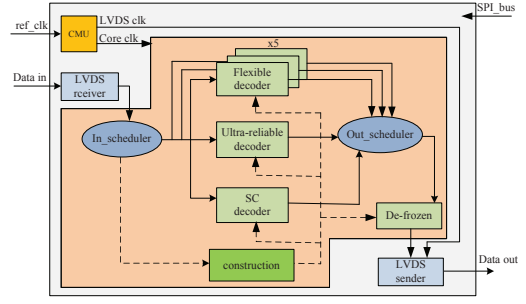} 
\caption{The architecture of the chip.}
\label{top architecture}
\end{figure}

The overview of our ASIC design is shown in Fig. \ref{top architecture}. It mainly comprises six units: SC decoder, flexible decoder, ultra-reliable decoder, de-frozen unit, code construction unit and scheduler. Five flexible decoders are integrated in the chip to achieve high throughput. The flexible decoder and ultra-reliable decoder can be configured as SCL, CA-SCL or PC-SCL during runtime. The code construction unit generates frozen bit set for all the three decoders. This can avoid transmission of frozen bit set and support any code rate. The de-frozen unit is responsible to remove frozen bits in source vector. 
The data-flow is managed by input-scheduler and out-scheduler.
\par There are four clock domains in the chip. All the six units are clocked by ``core clk", which is generated by clock management unit (CMU). The CMU also generates ``LVDS clk" for LVDS(Low Voltage Differential Signaling) sender. LVDS receiver and SPI(Serial Peripheral Interface) bus are clocked by external input clock.
\par The LLRs are represented in sign-and-magnitude form as in \cite{SC_ASIC}.
We denote  $Q_i,Q_c$ as the number of bits to represent internal LLR and channel LLR. In our ASIC design, we set $Q_c=6$ for all decoders.
We denote  $Q_{sort},Q_{PM}$ as the number of bits to represent PM in metric sorter and PM in memory, respectively.
After sorting, the minimum PM will be subtracted from the PM of all list paths and the quantization of PM will be reduced from $Q_{sort}$ to $Q_{PM}$. Therefore, $Q_{PM}$ of our ASIC is smaller than that of \cite{polarbear}.
\subsection{Flexible decoder}
\begin{figure}
\centering
\includegraphics[width=2.95in]{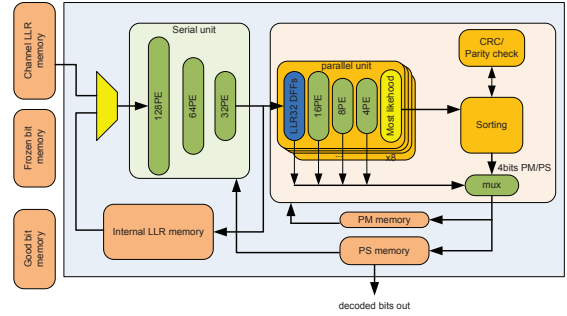} 
\caption{The architecture of flexible decoder.}
\label{scl8 architecture}
\end{figure}

Flexible decoder supports variable list size and code length, with upper limit $L_{max}=8$ and $N_{max}=2^{14}$. The architecture of flexible decoder is shown in Fig. \ref{scl8 architecture}. Channel LLR memory stories the received channel LLRs. Internal LLR memory stories the LLRs generated during decoding process. Good bits are information bits with higher reliability. They are stored and used to reduce path splitting as \cite{goodbit}. In this decoder, we set $Q_i=6$, $Q_{sort}=7$, and $Q_{PM}=6$ to preserve the same error performance as a floating-point decoder.
\par PEs are capable of performing f and g function.
If the parallelism of f/g node is larger than $16$, LLR processing is executed in the serial unit. Otherwise, it is executed in the parallel unit. 4-bit is decoded simultaneously in parallel unit by employing multi-bit decision \cite{6920050}. Up to $32$ rate zero nodes \cite{SSC} and rate one nodes \cite{SSC} in which all bits are good bits are also decoded simultaneously in parallel unit. Moreover, decoding starts from the first non-frozen bit as \cite{polarbear}.
\par We propose optimization techniques to improve throughput. They are employed in our decodes and presented below.
\subsubsection{LLR Memory Reduction}In a decoder chip, the internal LLR memory takes up most of the total core area. The $L_{out}$ in each stage should be stored and will be reused as shown in Fig. \ref{decoding_graph}. The memory size for internal LLR is
\begin{equation}\label{mem_cal_normal}
MEM=L \times Q_i \times \sum_{i=0}^{n-1} 2^i = LQ_i(N-1) .
\end{equation}
\par In a flexible decoder, we only save internal LLR for every three neighboring stages. The LLRs between these stages can be re-calculated on the fly from the stored LLRs.
Therefore, the memory size for internal LLR is reduced to
\begin{equation} \label{mem_cal}
MEM=L \times Q_i \times \sum_{i=n/3-3}^{n/3-1} 2^{3i} \approx 0.14\times LQ_i(N-1) .
\end{equation}
We can see that almost $86\%$ of internal LLR memory is reduced. To compensate the latency introduced by LLR re-calculation, more PEs are utilized.
\subsubsection{Serial List Processing} We propose a serial list processing architecture, which executes the LLR processing of different list in serial. As far as we know, all the hardware architecture in literature \cite{parallel_Architecture,6920050,7001058,fast_and_flexible} contains $L$ SC decoder cores and execute the LLR processing of different list paths in parallel. Due to LLR exchange among paths, this architecture requires a crossbar of LLR. The crossbar contains $L$ $L$-to-$1$ multiplexers with complexity growing proportional to $L^2$. 
\par Our serial list processing architecture executes LLR of different list one by one, thus does not require a crossbar. The LLR exchange among lists can be implemented by exchanging the address of memory. Compared to parallel architecture, serial architecture introduces no extra latency when PE quantity is the same. However, the complexity is reduced and the timing performance will significantly improve especially when list size and PE quantity is large.
\par At stage $t$, only $2^t$ f/g functions need to be executed as shown in Fig. \ref{decoding_graph}. Therefore, the large number of PEs in serial unit can not be fully used when $t<=4$. For these stages, we apply parallel unit to decrease the latency.
\subsubsection{Double-Package Mode}
Due to the serial nature of SC decoding, the PEs are idle during PM sorting period. The sorting latency is comparable with f/g execution latency when list size is large and multi-bit decoding is used. To improve the utilization ratio of PEs, double-package mode is applied when two packages are decoded simultaneously. When the PM of a package is sorting, the PEs are utilized to execute f/g function for the other package.
\par We define the decoding time as $T$ when only one package is decoded in the decoder. It requires less than $1.3*T$ to decode two packages under double-package mode. This can improve the throughput by $54\%$.
\subsubsection{Decoded-bit Recovery} In general, independent memory for $u$ and PS are required in decoder. Their sizes are both $LN$ bits\cite{parallel_Architecture}. However, the memory for $u$ is not necessary since $u$ is only required when decoded-bits are sent out.
\par Proposition $1$: For SCL decoder, $u$ can be recovered from PS after all bits are decoded.
\par Proof:
We denote $\hat{S}_t$ as the stored PS vector of stage $t$.
Since the serial nature of SC, only $2^t$ elements of $\hat{S}_t$ will be update simultaneously and need to be stored.
After all bits are decoded, the stored PS locates at the right lower triangle of decoding graph (e.g.,the green PS in Fig. \ref{decoding_graph}). We can infer that the final stored
\begin{equation}
\hat{S}_t=u_{N-2^{(t+1)}}^{N-2^t-1} \cdot G_t,
\end{equation}
where $G_t = F ^ {\otimes t}$. A characteristic of generator matrix G is $G=G^{-1}$. So, we can induce that
\begin{equation} \label{U_recovery}
u_{N-2^{(t+1)}}^{N-2^t-1}=\hat{S}_t \cdot G_t^{-1}=\hat{S}_t \cdot G_t .
\end{equation}
It can be seen as a polar encoding on $\hat{S}_t$. According to (\ref{U_recovery}), $u_0^{N-2}$ can be obtained. $u_{N-1}$ can be obtained when the last bit is decoded. Thus, $u$ can be recovered after all bits are decoded.
\par The polar encoding can be implemented by bitwise XOR. It takes much less chip area that $LN$ bits memory.
Furthermore, the encoding on PS can be executed in parallel with decoding of next package, has no effect on throughput.
Therefore, we use $LN$ bits memory to save PS and $N$ bits memory to save recovered $u$. Compared with general method, we can save $(L-1)N$ bits memory.
\subsection{SC Decoder}
The SC decoder is a simplified version of flexible decoder without path metric management. The SC decoder exploits SSC decoding algorithm \cite{SSC} and supports code length $N=2^{15}$. Due to the long code length, we set $Q_i=7$ to avoid error-correction performance loss.
\subsection{Ultra-reliable Decoder}
Aiming at $L=32$, we design an ultra-reliable decoder based on flexible decoder.
In the decoder, $Q_{sort}$ and $Q_{PM}$ is the same as those in the flexible decoder,
$Q_i=7$ at stage $0$ and $Q_i=6$ at other stages.
The main differences between ultra-reliable decoder and flexible decoder are shown below:
\subsubsection{Serial List Processing}There is a semi-parallel unit besides serial unit and parallel unit in ultra-reliable decoder. The number of PE, subjecting to the number of dependent nodes, can not be added arbitrarily to decrease the latency. Due to the large list size and the number of PE, the latency of serial processing deceases the throughput significantly at some stages.
Therefore, four list paths of stage $4\sim3$ are executed in parallel. These stages are called semi-parallel unit.
\subsubsection{LLR Memory Reduction}
Larger list size requires larger memory size for internal LLR  according to (\ref{mem_cal_normal}).
Therefore, we save internal LLR for every $4$ neighboring stages to save more memory in ultra-reliable decoder. However, four LLR copies of stage $5$ need to be stored for supporting semi-parallel unit and increasing the throughput. In total, almost $87\%$ of internal LLR memory are reduced compared with (\ref{mem_cal_normal}).
\subsubsection{Multi-bit Parallel Processing}Multi-bit decision is also adopted in ultra-reliable decoder. However, only 2-bit and up to $4$ rate $0/1$ nodes can be decoded simultaneously. 
\subsubsection{Double-Package Mode} Double-package mode is not supported to save memory.

\section{implementation Results and Measurement}
\begin{figure}
\centering
\includegraphics[width=3in]{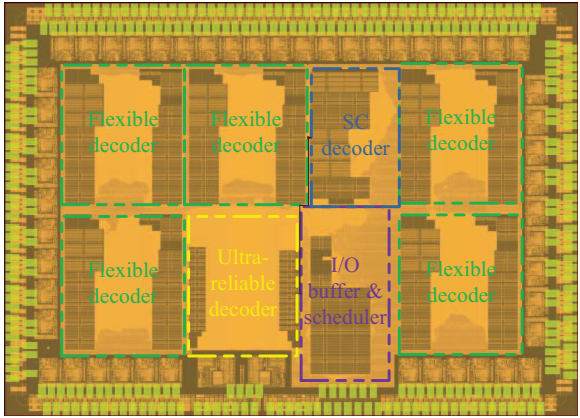} 
\caption{The micrograph of the decoder ASIC.}
\label{micrograph}
\end{figure}

\begin{figure}
\centering
\includegraphics[width=1.5in]{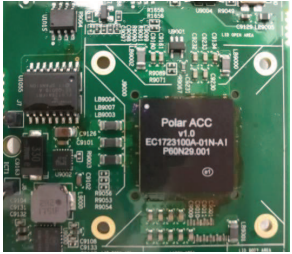} 
\caption{The photograph of the decoder ASIC.}
\label{graph}
\end{figure}

This ASIC is fabricated in a 16nm TSMC FinFET technology. The chip area is $6mm^2$ with $f_{clk}=1Ghz$, where $f_{clk}$ is the highest frequency of ``core clk". the micrograph and photograph of the chip is shown in Fig.\ref{micrograph} and Fig.\ref{graph}.

\subsection{Measurement Setup}
\par To test the decoder ASIC, we design a printed circuit board (PCB) which integrates the decoder ASIC and a Xilinx xc7vx690t FPGA. The PCB can be inserted into the PCIE slot of a computer. Test data is generated on the computer, and send to FPGA throughput PCIE. The FPGA acts as a bridge between computer and decoder ASIC.
\subsection{Error-Correction Performance and Throughput}
\begin{figure*}
\centering
\includegraphics[width=6.75in]{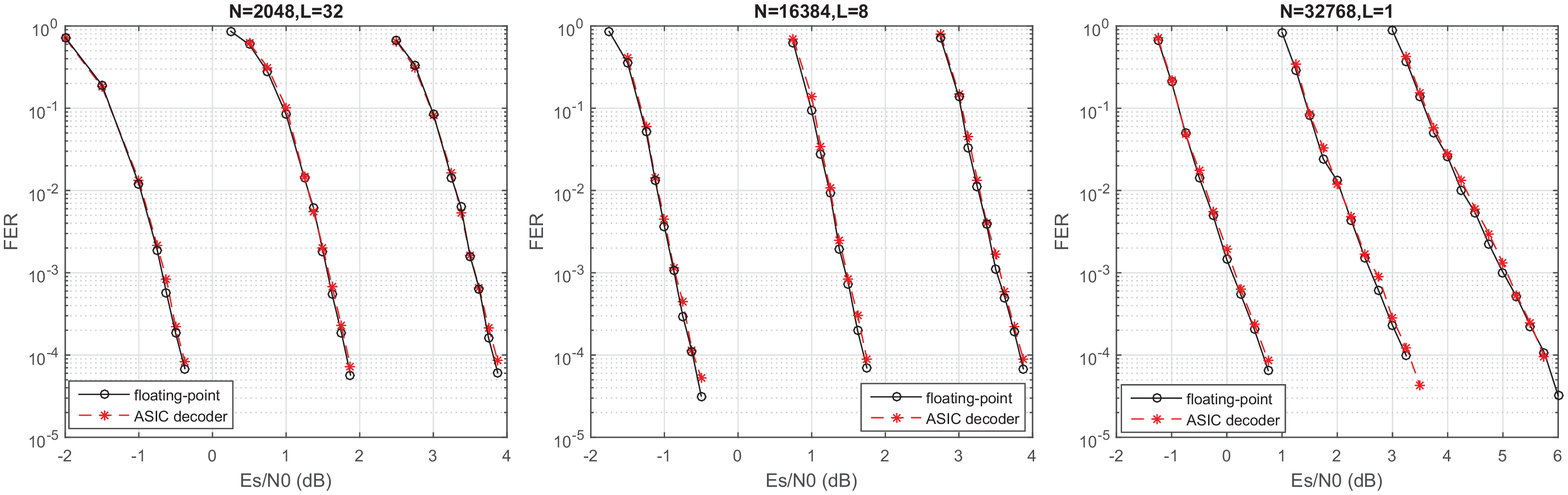} 
\caption{The error-correction performance of the three decoders under code rates $R \in [1/3,1/2,2/3]$.}
\label{bler}
\end{figure*}

The frame error rate (FER) under various list sizes, code lengths and code rates are tested by the designed PCB with $24$ bits CRC, and plotted in Fig.\ref{bler}. The codewords are randomly generated, modulated with quadrature phase-shift keying (QPSK) and transmitted over an additive white Gaussian noise channel. As a reference, the floating-point results are also plotted in Fig.\ref{bler}. It can be seen that quantization incurs performance loss less than 0.1dB.
\par The measured throughputs are summarized in Table. \ref{table_throughput}.
The throughput (T/P) is defined by 
\begin{equation}
T/P=k  \times f_{clk}/T ,
\end{equation}
where $T$ is the decoding latency. The highest throughput is $5.164$ Gbps when code rate $R=8/9$. The throughput of flexible decoder is even higher than the SC decoder since $5$ flexible decoder cores are implemented. In terms of area efficiency, SC decoder is the highest one in the three decoder.
Due to the large list size, the throughput of ultra-reliable decoder is much lower than the other two decoders.
 \begin{table}[!t]
    \begin{threeparttable}
    \caption{measured throughput}
    \label{table_throughput}
    \centering
    \begin{tabular}{|l|l|l|l|l|l|l|}
    \hline
    \diagbox{condition}{T/P (Mbps)}{code rate}   & 1/4  & 1/2    & 2/3     & 3/4  & 8/9 \\
    \hline
      $L=1$,$N=2^{15}$         & 1649    &2599        &3351       &3821  &4786 \\
       \hline
       $L=8$,$N=2^{14}$           & 1750      &2968      &3777         &4245   &5164   \\
       \hline
      $L=32$,$N=2^{11}$           & 33      &54         &70      &75  &91  \\
    \hline
    \end{tabular}
      \end{threeparttable}
    \end{table}
\subsection{Comparison With State-of-the-Art Fabricated ASICs}
The comparison with state-of-the-art fabricated ASICs is shown in Table \ref{table_example}. Our SC decoder supports $N=2^{15}$, but SC decoder in  \cite{polarbear} \cite{SC_ASIC} only supports $N=2^{10}$. Therefore, it is hard to give a precise comparison for these SC decoders in the table.
The flexible decoder and \cite{polarbear} run at the same code rate and length, but the former runs with larger list size $(L=8)$ and can support larger code length. Even though, the area efficiency of flexible decoder is much higher than the scaled result of \cite{polarbear}. As for ultra-reliable decoder, no fabricated ASIC decoder with $L=32$ has been reported in literature.
\begin{table*}[!t]
    \begin{threeparttable}
    \caption{comparison with state-of-the-art fabricated ASICs}
    \label{table_example}
    \centering
    \begin{tabular}{lllllllll}
    \hline
    implementation & SC decoder  & SC decoder & flexible decoder &ultra-reliable decoder &\cite{polarbear} & \cite{polarbear} &\cite{SC_ASIC} &\cite{6858413}\\
    \hline
    algorithm               & SC    & SC         &  SCL(L=8)  & SCL(L=32) &SC         &SCL(L=4)     &SC      &BP(15 iter)\\ 
    code length             & 32768 & 32768      &  1024      & 2048      &1024       &1024         &1024      &1024 \\
    code rate               & 1/2   & 869/1024    &  1/2      & 1/2       &869/1024     &1/2         &1/2      &1/2 \\
    technology              &  16nm  &  16nm      &16nm     & 16nm       &28nm  &28nm     &180nm   &65nm\\
    supply($V$)             & 0.9    & 0.9       & 0.9     & 0.9       &0.9     &0.9     &1.3    &1.0\\
    Frequency($MHz$)        & 1000   & 1000     & 1000  & 1000       &452     &308    &150    &300\\
    T/P (Mbps)              & 2599   & 4442    & 3241    & 54      &7836  &65.5  &49.0   &1024\tnote{(2)}\\
    area($mm^2$)            & 0.35   & 0.35    & 2.27    & 0.43      &0.35    &0.44   &1.71   &1.48\\
    area Eff. ($Mbps/mm^2$) & 7426   & 12691    & 1428    & 126      &22389    &148     &28.7    &692\\
    \hline
     \multicolumn{8}{l}{Normalized for 16nm\tnote{(1)}}  \\
    \hline
    T/P (Mbps)             & 2599          & 4442   & 3241   &54       &13713   &115      &551    &4160\tnote{(2)}\\
    area($mm^2$)           & 0.35          & 0.35   & 2.27   &0.43       &0.114    &0.144   &0.0135  &0.0897\\
    area Eff. ($Mbps/mm^2$) & 7426        & 12691   & 1428   &126       &120289    &793    &40815   & 46377\\
     \hline
    \end{tabular}
    \begin{tablenotes}
        \footnotesize
        \item [1] Area is scaled as $\lambda^2$, frequency as $1/\lambda$, where $\lambda$ is the technology feature size.
        \item [2] The throughput is scaled to worst case.
    \end{tablenotes}
      \end{threeparttable}
    \end{table*}
\par To further evaluate our architecture, we present the synthesis result of the flexible decoder and state-of-the-art decoders in Table \ref{table_sync}. We re-synthesize one flexible decoder and set $N_{max}=1024$ for fair comparison. The  $f_{clk}$ increases to $1.1GHz$.
The scaled result shows that the flexible decoder outperforms state-of-the-art decoders in terms of area efficiency.
 \begin{table}
    \begin{threeparttable}
    \caption{comparison of synthesis results for $N=1024$}
    \label{table_sync}
    \centering
    \begin{tabular}{llllll}
    \hline
    implementation   & This work  &\cite{fast_and_flexible} & \cite{7337462} &\cite{7114328} \\
    \hline
    list size           &  8          &8           &8         &8      \\ 
    technology          &16nm       &65nm         &90nm      &90nm   \\
    Frequency($MHz$)      & 1100    &722        &289       &637 \\
    T/P (Mbps)           & 713       &599       &374         &123 \\
    area($mm^2$)             & 0.06   &3.975    &7.22      &3.58 \\
    area Eff. ($Mbps/mm^2$)   & 11883    &151    &51      &34.4 \\
    \hline
     \multicolumn{5}{l}{Normalized for 16nm\tnote{(1)}}  \\
    \hline
    T/P (Mbps)                & 713     &2434   &2104        &692    \\
    area($mm^2$)               & 0.06    &0.241   &0.228    &0.113  \\
    area Eff. ($Mbps/mm^2$)   & 11883    &10124    &9228  &6124   \\
     \hline
    \end{tabular}
    \begin{tablenotes}
        \footnotesize
        \item [1] Area is scaled as $\lambda^2$, frequency as $1/\lambda$, where $\lambda$ is the technology feature size.
    \end{tablenotes}
      \end{threeparttable}
    \end{table}
\section{Conclusion}
In this paper, we present an ASIC implementation of three SC-based decoders for polar code in a 16nm TSMC FinFET technology.
To our knowledge, this is the first ASIC implemented SCL decoder supporting $L>4$ and $N>1024$.
To improve throughput, several optimization techniques are proposed.
Measurement result shows that throughput of the SC decoder, flexible decoder and ultra-reliable decoder can achieve up to $5.16$Gbps, $4.79$Gbps and $91$Mbps, respectively. Compared with fabricated AISC decoder and synthesized decoder in literature, the flexible decoder achieves higher area efficiency.

%
%



%
%

\bibliographystyle{IEEEtran}
\bibliography{IEEEabrv,polar_6858413CA_polarparallel_Architectureparallel_Architecturechip}

\end{document}